\documentclass[sigconf]{acmart}

\AtBeginDocument{%
  \providecommand\BibTeX{{%
    \normalfont B\kern-0.5em{\scshape i\kern-0.25em b}\kern-0.8em\TeX}}}

\setcopyright{acmcopyright}
\copyrightyear{2022}
\acmYear{2022}
\acmDOI{10.1145/3511047.3537657}

\acmConference[UMAP '22 Adjunct]{Make sure to enter the correct
  conference title from your rights confirmation emai}{July 04--07,
  2022}{Barcelona, Spain}

\acmBooktitle{Adjunct Proceedings of the 30th ACM Conference on User Modeling, Adaptation and Personalization (UMAP '22 Adjunct), July 04–07, 2022, Barcelona, Spain}
 
\acmPrice{15.00}
\acmISBN{978-1-4503-9232-7/22/07}

\begin{document}

\title{Using user's local context to support local news
}

\author{Payam Pourashraf}
\affiliation{%
  \institution{DePaul University}
  \streetaddress{243 S Wabash Ave}
  \city{Chicago}
  \state{IL}
  \country{USA}
  \postcode{60604}}
\email{ppourash@depaul.edu}

\author{Bamshad Mobasher}
\affiliation{%
  \institution{DePaul University}
  \streetaddress{243 S Wabash Ave}
  \city{Chicago}
  \state{IL}
  \country{USA}
  \postcode{60604}}
\email{mobasher@cs.depaul.edu}

\begin{abstract}
American local newspapers have been experiencing a large loss of reader retention and business within the past 15 years due to the proliferation of online news sources. Local media companies are starting to shift from an advertising-supported business model to one based on subscriptions to mitigate this problem. With this subscription model, there is a need to increase user engagement and personalization, and recommender systems are one way for these news companies to accomplish this goal. However, using standard modeling approaches that focus on users' global preferences is not appropriate in this context because the local preferences of users exhibit some specific characteristics which do not necessarily match their long-term or global preferences in the news. Our research explores a localized session-based recommendation approach, using recommendations based on local news articles and articles pertaining to the different local news categories. Experiments performed on a news dataset from a local newspaper show that these local models, particularly certain categories of items, do indeed provide more accuracy and effectiveness for personalization which, in turn, may lead to more user engagement with local news content.
\end{abstract}



\maketitle

\section{Introduction}
Personalization is all about creating a fulfilling experience for an end-user. To achieve this, the system requires knowledge of past user preferences and interactions to produce adequate recommendations. The interaction between the user and system facilitates designs that maintain customer relationships for an extended period.

The purpose of user engagement and personalization is customer advocacy. The end goal of customer advocacy is to work with existing and new customers to grow loyalty and retention. Because personalization has been proven effective for big businesses, it is important to consider it for local news in their geographical location. American local news has been drastically losing readers and business due to ineffective advertisement placements and an increase in digital readers. Now, local newspapers are trying to switch to subscription-based models that function on personalization and user engagement. Studies have shown the internet has become a major source of news due to its constant availability, and about nine in ten adults in the US read news online \citep{raza2021news}. However, an essential issue in the online media domain is the lack of definition of the criteria for newsworthiness (\citep{shoemaker2006news} as cited in \citep{raza2021news}). Therefore, creating a recommender system on a digital app is one-way local newspapers can increase their customer advocacy. The goal of the recommender system is to improve user experience and user engagement by prompting appropriate news articles to newsreaders.

Personalized news applications tend to prioritize news items based on topics in global user preferences across all news categories and geographical locations, as well as based on the popularity and recency of news items. However, the focus on long-term high-level preferences may, in fact, result in taking attention away from local stories that help provide the distinguishing added value for local newspapers. 

For example, a local newsreader may not generally be a sports fan, and hence sports-related features may not play a major role in generating personalized content for that user. However, the user, like many other local readers, might have an intense interest in the local high school team. Similarly, local readers may be interested in reading about local crime, even though for most users, this may not be part of their global news preferences. In such situations, global user preferences cannot be used to generate recommendations to users interested in local news content. At the same time, purely localized models will not be able to represent the users' broader interests that are needed to generate recommendations on more general topics such as national politics, entertainment news, etc. Ideally, the recommender system must combine global preferences with localized models for specific content categories in order to provide useful recommendations at all levels and to help increase user engagement on content where there is intense local interest among readers. 

In this research, we explore a session-based modeling approach to the personalized recommendation by focusing on different topic categories both at the local as well as global preference levels. We perform an offline empirical evaluation of different scenarios based on recorded past data rather than online experiments on a live system such as a news app. These experiments aim to inform the design of a hybrid recommendation framework that will allow local news outlets to provide added value and enhance user experience at the local level. Local news companies are shifting to digital subscription-based models because system effectiveness can translate to user satisfaction; personalized local news apps can then lead to reader advocacy and reader retention. We believe more focus on users' local interests would lead to a better recommender system to boost reader retention.

Thus the main Research Questions (RQ) of this work are as follows:

\noindent RQ1 - How do localized recommendation models perform in regards to system accuracy compared to standard models that focus on users' global preferences for news?

\noindent RQ2 - What is the effect of leveraging different local news categories on the overall quality of news recommendations?

We answer these research questions through a series of experiments based on a local news dataset from the city of Syracuse. We hypothesized that considering the context on the item side, which in this case is locality, would increase news recommender system accuracy for certain categories of local news. Our experiments will show that (a) considering a localized model is indeed helpful to improve the effectiveness of personalization along all of the considered evaluation metrics and (b) that the approach of focusing on subcategories will allow the system to identify those local news categories where localized recommendations are most useful.

There are three ways that user experience is commonly evaluated in recommender systems, according to \citep{konstan2012recommender}:
 (i) by carrying out user studies where the subjects are given certain questionnaires during different stages of recommendations, 
(ii) by combining study on longitudinally logged data with the questionnaire-based user study (iii) by addressing other evaluation measures such as combining accuracy and beyond-accuracy measures in certain ways. We used the third approach to evaluate our work.

This paper is organized as follows: Section 2 reviews related works. In Section 3, we discuss our dataset, and baselines methods metrics we used for our empirical experiments. In Section 4, we present details of the experimental design and a discussion of our results. The paper ends with a conclusion and future works in Section 5.

\section{Related Work}

Throughout the past decade, well-known news agencies such as The New York Times, Washington Post, and BBC have created personalized news feeds for their subscribers based on their profile or the content information of the articles read by the users. Increasingly these online systems use personalization to cater to users' interests and drive user engagement\citep{spangher2015building, graff2015washington, raza2020survey}. 
Research in news recommendation has explored a variety of methods such as Content-Based Filtering \citep{capelle2012semantics}, Collaborative Filtering \citep{das2007google}, Hybrid approaches \citep{das2007google}, Deep Learning \citep{qin2020research, de2018chameleon} and Graph-based models \citep{symeonidis2020session}.

These recommender systems have been trying to increase user engagement by filtering large article spaces and by assisting users in finding articles relevant to their interests through customization or personalization. \citep{jawaheer2014modeling}.

Recently, session-based recommendation algorithms in the news domain have been shown to be effective in recommending news items, in part because of the importance of factors such as recency and freshness in the news domain \citep{jannach2020research}. The goal is to consider the latest user session in session-based approaches and neglect or discount long-term user preferences \citep{jannach2020research}. In \citep{de2018chameleon}, the authors have proposed a deep learning meta-architecture named Chameleon on session-based news recommendations, which focuses solely on short-term users' preferences. The architecture includes a convolutional neural network for extracting textual features and a Long Short-Term Memory (LSTM) layer to model the sequence of clicked items in ongoing user sessions. 

Some efforts have recently considered both long-term and short-term user preferences in the recommendation system \citep{liu2018stamp, villatel2018recurrent, sun2020go}. In \citep{symeonidis2020session}, authors applied random walks with restart strategy and different time windows on heterogeneous graph networks to capture both users' long-term and short-term preferences. They applied their approach onto different news datasets and showed the efficiency of their approach.

In addition to these systems, there has also been research conducted around applying context to recommender systems in order to generate better recommendations. Context factors may include the time a user reads an article, the general location of the user, or the specific order of articles read by a user. An example is music recommendation systems that use different contexts, situations, and musical preferences of users to create a more accurate recommendation algorithm \citep{kaminskas2012contextual}. In the domain of news recommendation, the Chinese news recommender system CROWN \citep{wang2015crown}, for example, uses contextual user data such as time and location to increase the accuracy of recommendations and enhance user engagement. This real-time news recommender system uses contextual information, such as time, day of the week, the user's device, and a combination of different algorithms to account for the changes to user preferences during the day. 

While there have been many experiments with recommender systems in a broader news context, we found a research gap in using recommender systems for local news outlets in order to increase their reader retention rates.

We also wanted to explore using locality as an item-based context since previous research mostly focused on user-based context. However, locality in general does not necessarily lead to effective recommendation, even for local news sources. Readers of local news sources are interested in both local news as well content in broader non-local context. Furthermore, some local news categories  require specific models for recommendation due unique characteristics of these news categories and how users interact with them.

\section{Experimental Methodology}

Our approach involves considering local preferences in creating session-based news recommendation models that hopefully translate to increased user engagement. To conduct our empirical study, we collected and processed the data from a local news source. We designed our experiments based on the data to compare the localized models with standardized models and identify the effects of different news categories on overall recommendations.

\subsection{Dataset}

\begin{figure*}[h]
  \includegraphics[width=\linewidth,height=12cm]{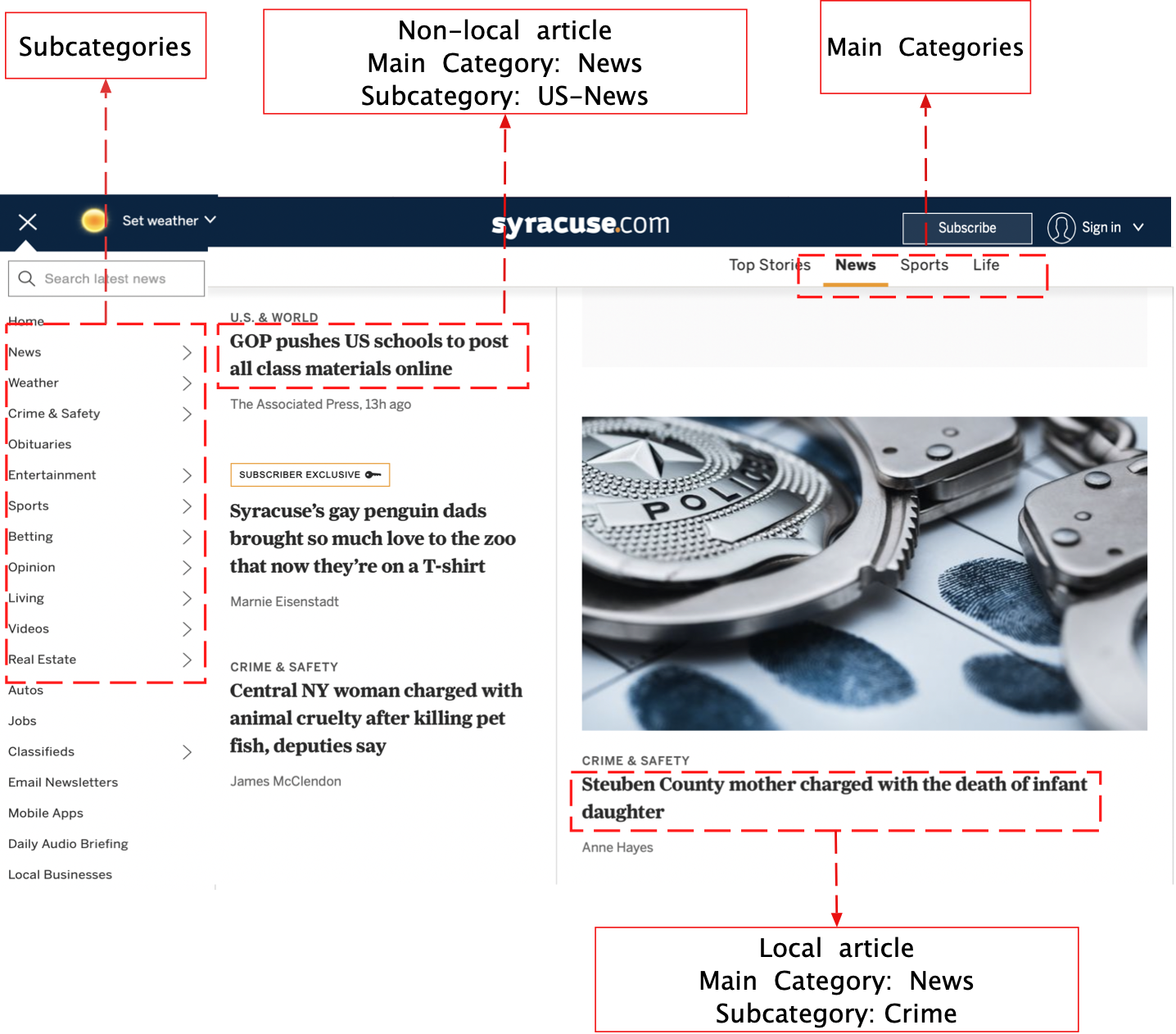}
  \caption{An example homepage of Syracuse news and two example news articles on it.}
  \label{fig: CH1}
\end{figure*}

The data from this research is roughly ten years of subscriber and user-interactions data from Syracuse News \footnote{https://www.syracuse.com}, a local newspaper in Syracuse, New York. The dataset has different content, behavioral, contextual, and session information. We organized this data using two categories of tags, one based on the articles' main types and one based on the URL information. The main categories that we used were news, sports, and life \& culture. Based on the URL, the second category is more specific and sorts the article into one of the subcategories. We've selected 4 months of all available user sessions for our experiments: December 1st, 2018 to March 31st, 2019. We elected to only focus on 4 months of data to account for the fact that this is a session-based model that would not need to be trained long-term. In a pre-processing step, we set a tag based on locality. Using a semi-automatic system, we used keywords that would define an article as global (i.e., US Government, Celebrity news, Lottery...) or local (i.e., Restaurants, Local sports teams, Local crime, Weather reports, State news, etc...). For example, the article titled "US government quietly spends millions to guard confederate cemeteries." was tagged as non-local, with "US Government" being the keywords used.
In contrast, the article titled "On this date: Binghamton 'evacuated' in nation's largest civil defense drill in 1957" was tagged as local since Binghamton is a New York county. For our dataset, we used Syracuse's city, so local news would include any articles that discussed New York state, New York counties, or cities of New York, in addition to global articles. Figure ~\ref{fig: CH1} shows how a Syracuse news source includes local and national news stories. The news source contains the top news categories and the list of various specific subcategories that the stories fall under.

\begin{figure*}[h]
  \includegraphics[width=\textwidth,height=5.5cm]{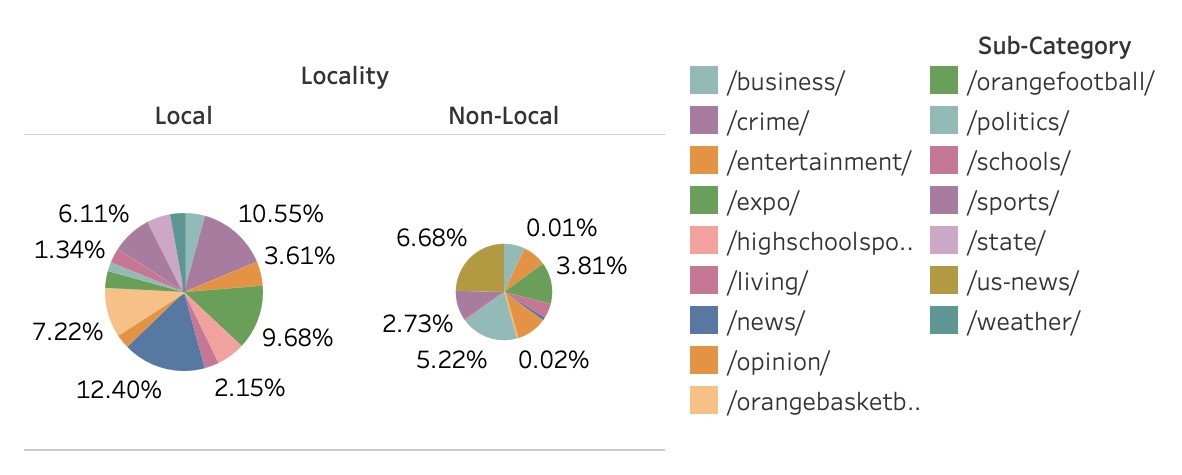}
  \caption{Distribution of the Syracuse dataset based on the sub categories (sports, news, and life \& culture)}
  \label{fig: CH2}
\end{figure*}

While many articles were categorized based on their headline, we manually read through a select number of articles to assign a tag accurately. The data was organized into sessions with a length of 1 day and sorted by their timestamp. Sessions with only one interaction were excluded since they are not helpful in our session-based recommender systems study. The results and statistics of the pre-processed datasets are shown in figure~\ref{fig: CH2} and tables~\ref{tab:freq}-~\ref{tab:commands}. In the first table, we can see we have 2826 non-local articles and 8145 local articles.

\begin{table}
  \caption{Detailed statistics of the Syracuse dataset.}
  \label{tab:freq}
  \begin{tabular}{cc}
    \toprule
    Period&12/1/2018 - 3/30/2019\\
    \midrule
    \texttt{\#articles} & 10971\\
    \texttt{\#local articles} & 8145\\
    \texttt{\#non-local articles} & 2826\\
    \texttt{\#sessions} & 60934\\
  \bottomrule
\end{tabular}
\end{table}

\begin{table}
  \caption{Breakdown of the dataset based on the main categories (sports, news, and life \& culture.)}
  \label{tab:commands}
  \begin{tabular}{cccc}
    \toprule
    Locality & News & Sports & Life \& Culture\\
    \midrule
   Local Articles& 34.02\% & 30.35\% & 11.43\%\\
   Non-local Articles & 13.92\% & 3.4\% & 6.89\% \\
    \bottomrule
  \end{tabular}
\end{table}

\subsection{Baseline Methods and Evaluation Metrics}
We applied the following baseline techniques for our experiments using the implementation from \citep{ludewig2018evaluation}: Association Rules (AR) \citep{agrawal1994fast,mobasher2001effective}, first-order Markov Chains (MC) \citep{norris1998markov}, Sequential Rules (SR) \citep{ludewig2018evaluation}, and Session-based KNN (SKNN) \citep{jannach2017recurrent}. We did not use the hybrid approach because the only content information we had from articles was the headline and URL. We had minimal contextual information surrounding the users, only looking at their location and time stamps. The deep learning approach needs a larger data set, so we neglect to report the results of that approach. We elected to use the so-mentioned simple baseline methods because they have mostly outperformed the more complicated approaches \citep{ludewig2018evaluation}. 

\begin{itemize}
\item {\verb|Association Rules (AR)|}: AR was based on the methodology proposed by \citep{agrawal1994fast} and discussed in \citep{mobasher2001effective}, which was the first application of association rule mining in the domain of recommender systems. This utilizes clickstream data to formulate a model that effectively generates web news personalization. We used reference \citep{ludewig2018evaluation} adaptation of AR, which is a simplified version of association with the rules of size two. When an item appears after another item in a session, the rules are being created regardless of the occurrence gap. No minimum support or confidence thresholds were applied following our utilization of \citep{ludewig2018evaluation}.
\item {\verb|Markov Chains (MC)|}: In MC, the rules are extracted from a first-order Markov Chain, and the sequence in the data is being considered \citep{norris1998markov}. Each item's score in a given session is calculated by counting the frequency of users who viewed other items right after viewing that item. The MC describes the transition probability between two subsequent articles in a single session. It calculates how often a reader views a specific article after viewing another one.
\item {\verb|Sequential Rules (SR)|}: SR is similar to AR and MC. It considers the items' order, but unlike MC, with a minimum of one step distance between the two items. The rule is then getting a weight associated with the item-gap inverse (one over the number of steps between the two items). From the currently viewed item in one session, we set the maximum number of clicks to step back to 10. Another important hyper-parameter was the decay function, which measures the distance between two clicks set to a linear function \citep{ludewig2018evaluation}.
\item {\verb|Session-based KNN (SKNN)|}: SKNN compares the entire current session with the training data's past sessions to determine the recommended items. Rather than considering only the last event in the current session, as in ItemKNN, the SKNN method compares the entire current session with the training data's neighboring sessions. After trying several values for k, we decided to use k = 20 because it suited our experiment the best. For the session similarity measure, we used cosine similarity, which was used to find the k most similar past sessions for any given session \citep{jannach2017recurrent}.
\end{itemize}


There are two main evaluation methods: objective measures, such as accuracy and diversity, and subjective measures, such as reader satisfaction. This study focused on objective measures since we were using offline experiments and did not have access to the users. To measure quality factors such as accuracy, we have selected a set of metrics such as Hit Rate (HR), NDCG, and Mean Reciprocal Rank (MRR)\citep{parra2013recommender, ozgobek2014survey}.
\begin{itemize}
\item {\verb|HR@K|}: Among the top K items (articles in our experiments) HR finds the rate of times in which relevant items are retrieved \citep{parra2013recommender, wu2021personalized}.
\item {\verb|NDCG@K|}: is a standard ranking metric that uses the graded relevance to rank each item (which commonly, an article is viewed as either relevant or not relevant by the other ranking metrics) \citep{parra2013recommender, wu2021personalized}.
\item {\verb|MRR@K|}: is a ranking metric that takes the average of the reciprocal ranks of the top K results for a sample of recommendation lists. MRR considers the rank of the item \citep{parra2013recommender, wu2021personalized}.
\end{itemize}

\section{Experimental Results and Analysis}

In this section, we present our experiments to address the research questions (RQ1 and RQ2). As noted earlier, the primary goal of these experiments is to determine the degree to which models trained on certain local news categories outperform recommendation models based on general and global user preferences. In this context, we also want to identify scenarios in which broader models based on both local and non-local preferences are more appropriate to provide the most effective recommendations. Our ultimate goal from this study is to gain insights into the best way to design hybrid news recommenders (combining both localized and general models) for local news outlets.

\subsection{Experimental Design}

Using our local news dataset (presented in section 3.1), we experimented with different session-based recommendation methods (see section 3.2) in three different scenarios to explore the effectiveness of localized models in a news recommendation system. 

In the first scenario, all articles (both local and non-local) were used in the training of the models as well as in the testing phase. In the second scenario, only local articles were used to learn localized models, and these models were also tested only on local news data. 
In the third scenario, models were learned based on all data, but were only tested on test data comprising of local news articles. We used HR@K, NDCG@k, MRR@K for respective datasets to measure the system's performance in each scenario. 

For training and testing splits, we used one single train-test split. The data was split in such a way that the sessions of all four months except those of the last ten days of the entire dataset were placed in the training set. We used the articles from the last ten days for testing. We report the results of applying this evaluation scheme using the Syracuse dataset. Sessions with only one interaction were excluded as we could not use them for the next-lick prediction.

In the initial experiment, we compared the performance of our baseline session-based recommender systems in the above three scenarios. It should be noted that our aim here is not to identify the best algorithm for making news recommendations. Instead our goal as to address RQ1, i.e., to determine how much more effective the recommendations would be when the algorithm focuses on local news. We compared the local and general models using several different algorithms to show that the results were not algorithm-specific. 

To address RQ2, we designed an experiment to determine the impact of "hyper-local" models where models were trained and/or tested on specific news categories and subcategories both at the local and the global levels. With these experiments, we hope to find effective models to recommend different local article categories. We also wanted to distinguish local news categories where these localized models are beneficial and those categories where the broader and global preference models are more appropriate.

So, in the next set of experiments, we implemented nine different scenarios to find the effectiveness of training models on all vs. training on local articles in various news categories. We chose the categories of sports, life \& culture, and news since they have the largest amount of articles compared to other categories. We created three scenarios with each of these categories to test the model effectiveness when trained on all articles (local and non-local) compared to when it's trained on just local articles (in different categories). The first scenario trained on all articles and tested on local sports articles, the second trained on all articles and tested on all sports, and the third trained on local articles and tested on local sports articles. We ran these three scenarios following the same protocol as the first experiment, using performance metrics such as HR@K, MRR@K, and NDCG@K. We also replicated these experimental scenarios in all three categories mentioned above.

\subsection{Main Results and Discussion}

\begin{table*}
\small
  \caption{Results on the test set of Syracuse Local and All datasets.}
  \label{tab:local-all}
  \begin{tabular}{lcc|cccccc}
    \toprule
     Method&  Training Set & Testing Set & HR@10& HR@20& MRR@10& MRR@20&NDCG@10& NDCG@20\\
    \midrule
   SKNN& All & All  & 0.4368 & 0.5460& 0.1288&  0.1366& 0.2566& 0.2852\\
   &  Local& Local & \textbf{0.4940}& \textbf{0.5938}& \textbf{0.1446}& \textbf{0.1514}& \textbf{0.2926} &\textbf{0.3183} \\
      &  All & Local & 0.4584& 0.5653 & 0.1378 & 0.1452 & 0.273 & 0.3008 \\

    \midrule
   MARKOV& All & All  &0.3276 & 0.3918 & 0.16096 & 0.1655 & 0.2295 & 0.2463\\
   &  Local& Local & \textbf{0.361} & \textbf{0.4441} & \textbf{0.1845} & \textbf{0.1904} & \textbf{0.2547} & \textbf{0.2764}\\
      &  All & Local & 0.3396 & 0.4109 & 0.1700 & 0.175 & 0.2393 & 0.25796 \\
    \midrule
   AR& All & All  & 0.4304 & 0.5117 & 0.2069 & 0.2128 & 0.2997 & 0.3213\\
   &  Local& Local & \textbf{0.4631} & \textbf{0.5486} & \textbf{0.2198} & \textbf{0.2266} & \textbf{0.31930} & \textbf{0.3437} \\
      &  All & Local & 0.456 & 0.5391 & 0.2147 & 0.22 & 0.3184 & 0.3382 \\
    \midrule
    SR& All & All  & 0.3918 & 0.471 & 0.1732 & 0.17892 & 0.259 & 0.2798\\
   &  Local& Local & \textbf{0.4156} & \textbf{0.5011} & \textbf{0.2090} & \textbf{0.2151} & \textbf{0.2969} & \textbf{0.3193} \\
      &  All & Local & 0.4133 & 0.4916 &  0.1873 & 0.1929 & 0.2766 & 0.2972\\
    \bottomrule
  \end{tabular}
\end{table*}

The results of our first experiment are depicted in Table ~\ref{tab:local-all}. These result show that, in general, recommendation models that are trained on local news articles tend to be more accurate when recommending local articles than recommenders that were trained on both local and non-local articles. 

Also, looking at the results across different recommendation algorithms, the session-based KNN model (SKNN) generally performed better than the other methods with respect to all three metrics. More importantly, the overall finding of the effectiveness of localized models is not affected by choice of the algorithm. For these reasons, in subsequent experiments, we only used SKNN to compare the performance of different models in the context of different scenarios. 

With the second set of experiments (shown in Table~\ref{tab:maincategory}), we found that, when recommending local articles, training on local articles yielded better personalization than training with all articles. This effect was observed across all of the scenarios involving local news categories. These findings support the conjecture that users' local news preferences may differ from their general preferences on news topics. For example, users may not be interested in general news articles related to the life \& culture category. However, specific local art and culture events are often the subject of intense interest on the part of local readers. Such articles may be recommended using localized recommendation models but may be omitted from recommendation lists otherwise.

We also observe a difference in the degree of improvement in recommendation effectiveness across categories. For example, there is a greater advantage in using localized models for the Local News Category than there is for the other two categories.In fact, it may be possible that the more general models trained on all news articles perform better than localized models in the case of some news categories. This indicates that simply training based on local news is not enough. The overall system must be calibrated carefully to take advantage of those localized models where there is a significant difference in user preferences between local news items and general news items. 

The localization effect mentioned above is even more evident when more granular categorization of local news articles is used. In order to find specific characteristics of users' local preferences, we decided to explore the main categories by breaking them down into subcategories. Referring to the 3.1 section, the article subcategories are based on URL information. For example, the URL of the article, which is titled “A closer look at Chevy's 2020 silverado hd pickup that will debut in February," is: "\url{https://www.syracuse.com/auto/2018/12/new_silverado_in_february_2019.html}". Based on this, the extracted subcategory is “/auto/” which is the first subdirectory of the URL. The main categories and subcategories are not mutually exclusive sets. For example, News and Life \& Culture are in two separate main categories but share the same /auto/ subcategory.

In each scenario, the main categories are broken down to probe how the top subcategories perform, which is shown in Table~\ref{tab:subcategory}). The performance using HitRate@20 shows significant variance across different subcategories depending on training sets (local vs. all). For example, in the subcategory of News/Politics, we can observe a higher performance when the model is trained on all articles. This is an example of the situation where localized models would be less effective than global models. In this case, the effect is likely due to the fact that among local readers, their news preferences related to politics center more around non-local and national issues.

\begin{table}
\small
  \caption{Results for different scenarios on various news categories using SKNN method.}
  \label{tab:maincategory}
  \begin{tabular}{cc|ccccc}
    \toprule
     Training Set & Testing Set &  HR@20&  MRR@20 &  NDCG@20\\
    \midrule
    All & All& 0.5460 & 0.1366 & 0.2852\\
   Local & Local&  \textbf{0.593} & \textbf{0.1514} & \textbf{0.3183}\\
 All & Local& 0.5653 & 0.1452 & 0.3008\\
    \midrule
    All & All Sports&  0.5410 & 0.1334 & 0.2862 \\
  Local & Local Sports& \textbf{0.5583} & \textbf{0.1397} & \textbf{0.2930} \\
All & Local Sports&  0.5329 & 0.1327 & 0.2839\\
    \midrule
   All & All Life\&Culture&  0.5797 &  0.1906 & 0.3478 \\
  Local& Local Life\&Culture& \textbf{0.6557} & 0.1885 & \textbf{ 0.3919}\\
All & Local Life\&Culture&  0.6393 & \textbf{0.2125} & 0.3865\\
    \midrule
   All & All News& 0.6460 & 0.1729 & 0.3553\\
  Local & Local News& \textbf{0.75} & \textbf{0.2015} & \textbf{0.4240}\\
All & Local News&  0.6979 & 0.1912 & 0.3909\\
    \bottomrule
  \end{tabular}
\end{table}

On the other hand, the results for the subcategory of Sport/Orange basketball (a local basketball team) shows the opposite effect, with significant increase in performance from the non-local models. It should be noted that additional experiments with larger datasets are needed to draw definitive conclusions. But, the results reasonably support the assertion that for some specific local news categories, it is important train localized recommendation models and incorporate then into a larger ensemble or hybrid system. 
.
\begin{table*}
\small
  \caption{Results on various news subcategories.}
  \label{tab:subcategory}
  \begin{tabular}{cc|c|c|c}
    \toprule
    Training Set & Testing Set & News/Crime& News/Politics& News/news\\
 \midrule
    Local& Local &0.25&0.1 & 0.127\\

    Local & Local News & \textbf{0.46}&0.238&0.173\\    

    All& All &0&0.189474&0.152\\    

    All& All News &0&\textbf{0.340}&\textbf{0.214}\\    
   \toprule
    &  & Sport/sports& Sport/Orange basketball&Sport/highschool sports\\
    \midrule
    Local&Local &0.095&0.596&0.32704\\
    Local&Local Sports &\textbf{0.1607}&\textbf{0.7622}&\textbf{0.3798}\\    
     All&All &0.117&0&0\\    
    All & All Sports &0.155&0&0\\ 
    \bottomrule
  \end{tabular}
\end{table*}

\section{Conclusion}

In this paper we proposed using localized models in session-based recommender system, particularly when recommending local news items where there is intense local interest and there is significant difference between users' global and local news preferences. These findings can inform the design of more effective local news recommender system that distinguish between certain categories of local and general news. 

We conducted multiple offline experiments on a local news dataset that showed how localized models in specific news categories performed better compared to models based on general news preferences. We did a case study that shows there is a distinction between the way users interact with local news vs. global news, and for some categories there is a significant difference in users' global and local preferences.

Moving forward, we plan to conduct additional experiments with datasets from other local news sources, possibly combining these datasets to allow for experiments with more granular category structures while reducing the possibility of over-fitting. We also plan to explore the effect of model localization on beyond accuracy metrics such as diversity and novelty. 

Furthermore, we plan to develop a general recommendation framework for local news that involves automatically identifying specific categories of news that benefit from localized models and to combine these localized models with other models based on general preferences into an integrated system. 

In order to truly understand the impact of localized models on recommendation effectiveness, eventually we must perform a user study involving a live experimental system. We intend to explore the creation of such a system, possibly as an extension of an existing online local news site, to conduct such a user study.


The ultimate goal of this and future efforts in this direction are to find ways to more effectively derive user engagement for local news outlets which in turn helps these outlets generate more subscription and advertising revenue. Such effective recommender systems may be among the tools that can help reverse the current trend toward the extinction of locally sources news.

\bibliographystyle{ACM-Reference-Format}
\bibliography{sample-base}

\end{document}